\documentclass[twocolumn,showkeys,showpacs,superscriptaddress,preprintnumbers,amsmath,amssymb,prl]{revtex4-1}
\usepackage{graphicx}
\usepackage{epstopdf}
\usepackage{dcolumn}
\usepackage{bm}
\usepackage{amsmath}
\usepackage{color}

\newcommand{\bl}[1]{\textbf{#1}}

\begin{document}

\title{Force and Mass Dynamics in Non-Newtonian Suspensions}

\author{Melody X. Lim}
\altaffiliation[Current address: ]{Department of Physics and James Franck Institute, The University of Chicago, 5720 S. Ellis Ave, Chicago, Illinois 60637, USA}
\affiliation{Department of Physics \& Center for Nonlinear and Complex Systems, Duke University, Durham, North Carolina 27708, USA}
\author{Jonathan Bar\'{e}s}
\altaffiliation[Current address: ]{LMGC, UMR 5508 CNRS-University Montpellier, 34095 Montpellier, France}
\affiliation{Department of Physics \& Center for Nonlinear and Complex Systems, Duke University, Durham, North Carolina 27708, USA}
\author{Hu Zheng}
\affiliation{Department of Physics \& Center for Nonlinear and Complex Systems, Duke University, Durham, North Carolina 27708, USA}
\author{Robert P. Behringer}
\affiliation{Department of Physics \& Center for Nonlinear and Complex Systems, Duke University, Durham, North Carolina 27708, USA}

\begin{abstract}

Above a certain solid fraction, dense granular suspensions in water
exhibit non-Newtonian behavior, including impact-activated
solidification. Although it has been suggested that solidification
depends on boundary interactions, quantitative experiments on the
boundary forces have not been reported. Using
high-speed video, tracer particles, and photoelastic boundaries, we determine the
impactor kinematics and the magnitude and timings of impactor-driven
events in the body and at the boundaries of cornstarch suspensions. We
observe mass shocks in the suspension during impact. The
shockfront dynamics are strongly correlated to those of the
intruder. However, the total momentum associated with this shock never
approaches the initial impactor momentum. We also observe a faster
second front, associated with the propagation of pressure to the
boundaries of the suspension. The two fronts depend differently on the
initial impactor speed, $v_0$, and the suspension packing fraction. The speed
of the pressure wave is at least an order of magnitude smaller than
(linear) ultrasound speeds obtained for much higher frequencies,
pointing to complex amplitude and frequency response of cornstarch
suspensions to compressive strains.

\end{abstract}
\date{\today}


\maketitle

Dense suspensions, such as cornstarch in water, provide a rich set of
phenomena, including complex non-Newtonian response to shear
\cite{Brady85,Wagner09,Brown10,Roche13,Wyart14,BrownJaeger14},
discontinuous shear thickening (DST) \cite{Wyart14}, and
impact-activated solidification (IAS), the focus of this study. Brown and Jaeger
\cite{BrownJaeger14} provide a good snapshot of the field. Cornstarch suspensions are also remarkable for striking behavior
  such as the formation under strong vertical vibration of
  holes and fingers~\cite{merkt04,vankann14}. Also, it is possible to run but not walk across
  a pool of suspension without sinking.  During IAS,
the suspension responds to a rapid impact with large normal
stresses. DST and IAS occur for packing fractions, $\phi$, that are
close to the jamming transition, i.e. where suspensions, granular
materials, etc. become solid-like and support finite stresses~\cite{Ohern03,Bi11,Seto13,Wyart14,Peters16} The details of the {\em
  dynamics} of impacts into suspensions is crucial to understanding
the nature of IAS, and its connection to a much broader range of
static and dynamic phenomena in suspensions and other particulate
systems. Frictional granular materials jam under shear strain
\cite{Bi11} for $\phi < \phi_{J-frictionless}$, and frictional effects
may play an important role in suspensions
\cite{Seto13,Wyart14,Peters16}. Impact experiments on dry frictional granular
systems \cite{Wildenberg13,Clark15,Clark_pre16} show shock-like
response, where again friction/no-friction matters
\cite{Clark15}. Experiments
\cite{Liu10,Jaeger12,Jaeger14,Brown14,Kann11} suggest that during IAS,
a dynamic jamming and unjamming process occurs: the suspension
temporarily solidifies above a critical impact velocity
\cite{Brown14}, and the force on the impactor depends on interactions
with the suspension boundary \cite{Brown14,Jaeger14}. There may also
be a connection between force propagation in impact experiments and
non-locality/cooperativity reported recently
\cite{Katgert10,Goyon08,Kamrin12}.

However, quantitative experiments on the forces experienced by the
boundaries of the suspension have not been reported, to our
knowledge. An important finding of this work is a fast pressure signal
that reaches the boundary before  a mass
shock, and carries the majority of the momentum. This signal may
provide insight into non-local response in other particulate systems.

Here, we measure the strain response within the suspension, and the
force response from the boundaries of the suspension, due to impact of
an intruder into a vertical channel of a water-cornstarch
suspension. We observe a transient solid within the suspension, with
dynamics that are strongly correlated with those of the impactor, and
a second impactor driven front, with different dynamics from the solid
front.

We correlate the dynamics of the fronts formed inside the suspension
and the forces on the suspension boundaries with the impactor
 dynamics using two separate sets of experiments. In both, we dropped
 a metal disc from varying heights into a cornstarch suspension with
 packing fractions~$0.38<\phi<0.48$. In most experiments, the suspension was enclosed in a rectangular
 acrylic channel ($h\times l\times w = 177 \times 138 \times 15$ mm),
 with $\sim 35$\% occupied by the gelatin boundary, whose shape is
 shown in Fig.~\ref{fig:photoelastic-time-intensity}A. The disk was
 guided by a chute located above the container. The disk had a
 diameter of 63.5mm, width 11mm, mass 291g, and had a 10mm hole in the
 center for tracking. We recorded impacts with a Photron FASTCAM
 SA5. We tracked the impactor using a circular Hough transform at each
 video frame, then numerically computed the velocity and acceleration
 of the impactor, filtering out noise with a low-pass filter
 (cut-off~$200$ Hz). We tracked the intruder position in both the
 polarized and unpolarized experiments, described below.

To gain access to the boundary stresses, (first experiments) we lined one
side of the container with gelatin, a good photoelastic material
\cite{kilcast84} that has a low friction coefficient with acrylic
($\sim 0.01$). The container plus suspension was placed between
crossed circular polarizers, yielding the photoelastic boundary
response. The apparatus was lit from behind using a halogen lamp with
diffuser hood. To capture the photoelastic response, we recorded video
at a frame rate of 42,000 frames per second.

The second experiments visualized the suspension flow field. The
suspension was mixed with tracer particles (black glitter, diameter
$\sim 0.25$mm). We recorded the tracer particle motion with front
lighting, and without crossed polarizers at 10,000 frames per
second. We used particle image velocimetry (PIV) to extract the
velocity field inside the suspension and to deduce the position of the
wavefront. We also characterized the flow of the material by finding
the difference between successive frames from direct high speed video,
producing a space-time plot of the movement of tracer particles in the
suspension.

In order to test whether there was significant coupling between the
Plexiglas faces and the cornstarch, we carried out a third
and limited study in which additional $2.5$~mm thick layers of gel
were placed between the Plexiglas and the cornstarch. As discussed in
supplementary information, the propagation speeds were not affected by
replacing the Plexiglas faces with soft faces.

We are therefore able to combine data for the timings of (i) impactor
dynamics events, (ii) photoelastic boundary events, and (iii) events
in the body of the suspension, providing insight into the physical
basis of the impact-activated solidification of the cornstarch
suspension.

\begin{figure}
\centering \includegraphics[width=0.71\columnwidth]{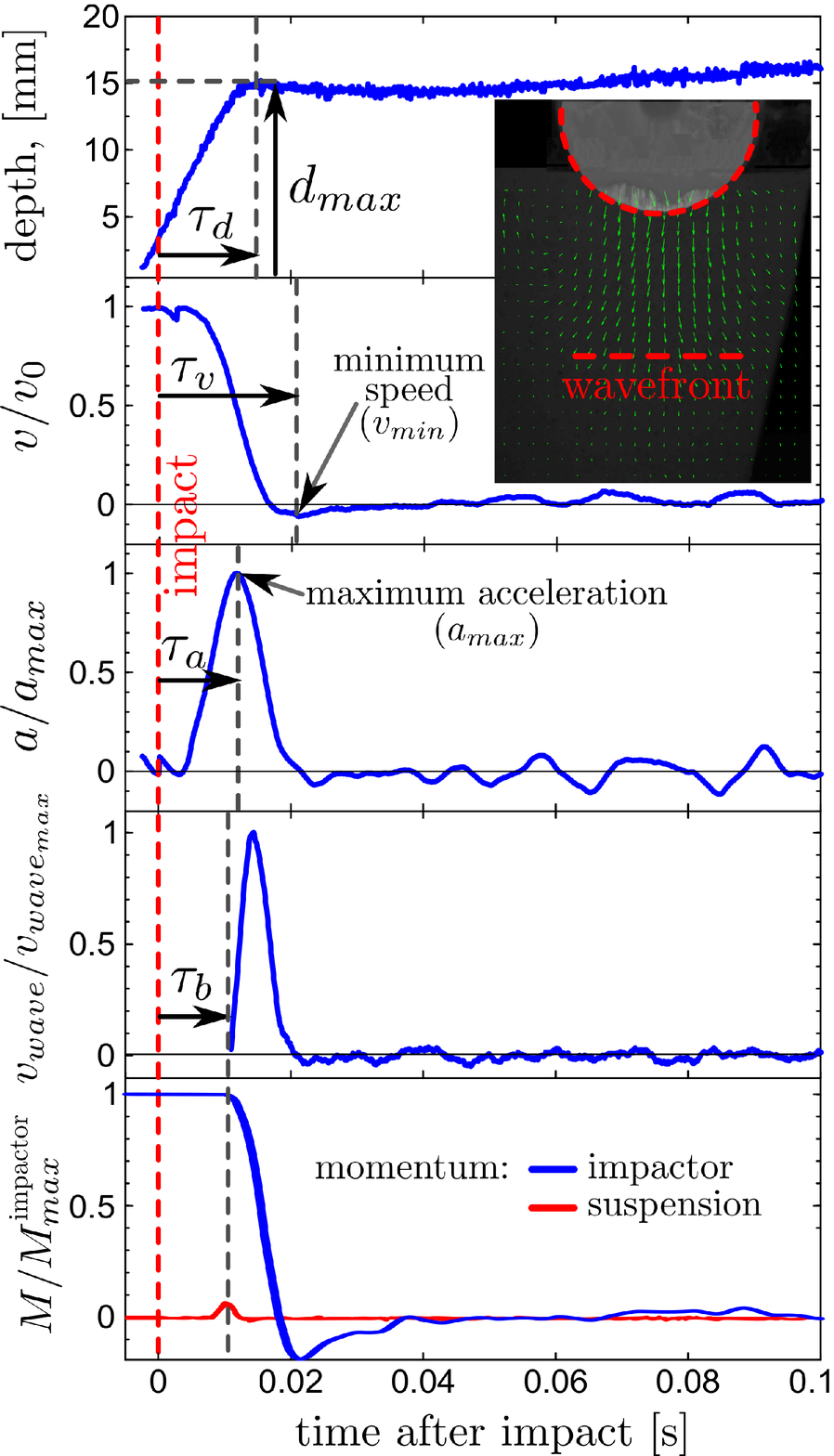}
\caption{(color online) Inset: An impact on the suspension with~$\phi=0.42$
  showing the position of the impactor (dashed circle) and the
  velocity field within the suspension (see supplementary material for
  video), derived using particle image velocimetry (PIV, green
  arrow). The wavefront position in each frame is extracted from the
  dashed line, which is numerically differentiated to give the front
  speed. Main panel: Time series are shown for (from top to bottom):
  the depth of impactor, the velocity of impactor normalized by impact
  velocity, $v/v_0$ ($v_0=1.9$m/s), the acceleration of the impactor
  normalized by maximum acceleration, $a/a_{max}$
  ($a_{max}=360$m/s$^2$), the speed of the mass shock within the
  suspension normalized by its maximum velocity,
  $v_{wave}/v_{wave_{max}}$ ($v_{wave_{max}}=1.5$m/s), and the
  momentum of the impactor and suspension normalized by the initial
  impactor momentum. Upon impact, the impactor rebounds from the
  surface of the suspension, as if colliding with an elastic solid,
  but also sinks slowly into the suspension after rebounding, as if
  into a viscous liquid. Note the
  well defined time-series of events after impact ($\tau_a$, $\tau_b$,
  $\tau_d$, $\tau_v$). Additionally, the momentum transferred to the
  suspension never approaches the initial impactor momentum (data shown for a different experiment).}
\label{fig:impactor-dynamics}
\end{figure}

\begin{figure}
\centering \includegraphics[width=0.75\columnwidth]{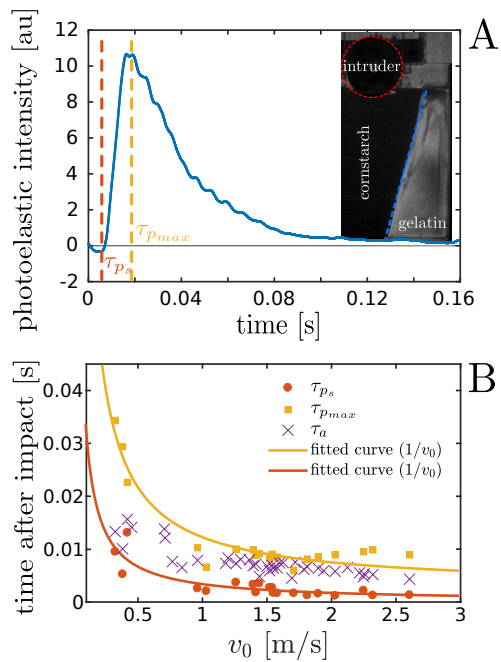}
\caption{(color online) Events in the boundary of the photoelastic
  material. Inset: An image of the photoelastic signal from the boundary
  during an impact with~$\phi=0.42$ (see supplementary material for video).  A: Total
  intensity of the signal from the photoelastic boundary as a function
  of time. B: The times for the two events of part B are plotted
  vs. initial impactor speed~$v_0$, along with fitted curves. For a more complete picture of events in the suspension, we also plot~$\tau_a$ on the same axes (crosses).}
\label{fig:photoelastic-time-intensity}
\end{figure}

We begin with the dynamics of the impactor. These events and their
timing are shown in Fig. \ref{fig:impactor-dynamics}. After striking
the surface of the suspension, the impactor settles into the
suspension for a short period of time. It then experiences a
significant upward normal force from the suspension, stopping the
impactor motion at a maximum depth,~$d_{max}$. This normal force lasts
for an extremely brief period of time, as seen in the rapid increase
in the intruder acceleration to~$a_{max}$. This causes the impactor to
rebound with a peak velocity~$v_{min}$, before settling into the
suspension at a much lower speed.

Beneath the impactor, the local suspension velocity just after impact
rises well above the background fluctuations, as in
Fig.~\ref{fig:impactor-dynamics} inset. The lower boundary of this
region moves downward with speed~$v_{wave}$ that is strongly
correlated with the impactor acceleration, producing a sharp peak in
the cross-correlation signal between the two (not shown), as well as a small delay between the time of maximum impactor acceleration,~$\tau_a$, and~$\tau_w$, the time of maximum~$v_{wave}$. This
suggests the formation of a solid mass beneath the impactor, which
might transmit internal stresses quickly, and moves at a uniform
velocity.

We estimate the total momentum transferred to the suspension over the
course of an impact by integrating over the velocity field shown in
Fig.~\ref{fig:impactor-dynamics} inset, multiplying each velocity
element by its associated volume and density, and assuming that the
observed motion at the surface is typical of the motion inside the
sample. The total momentum transferred to the suspension as a function
of time, along with the momentum of the impactor over the course of
the same impact, are shown in the bottom panel of
Fig.~\ref{fig:impactor-dynamics}. Strikingly, the momentum transferred
to the suspension never approaches the initial intruder momentum,
suggesting that the majority of the impactor momentum must be absorbed
by the apparatus, without appearing in the mass flow of the
suspension.

\begin{figure}
\centering \includegraphics[width=0.71 \columnwidth]{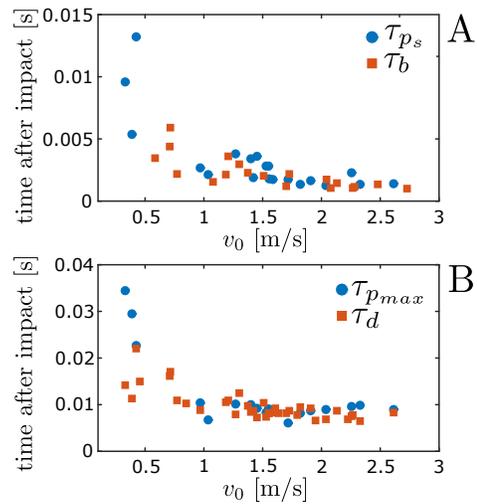}
\caption{(color online) Correlations between the timings of events in the suspension boundary, in the impactor, and in the motion of the solid mass within the suspension ($\phi=0.42$). A: The time at which the first signal at the boundary is received,~$\tau_{p_s}$, is the same as the time at which bulk motion is first observed beneath the impactor,~$\tau_b$. B: The time at which the stress on the boundary is maximal,~$\tau_{p_{max}}$, is the same as the time at which the impactor reaches its maximum depth,~$\tau_d$.}
\label{fig:impactor-photoelastic-timeseries}
\end{figure}

These results indicate several key timings in the suspension and
impactor dynamics for comparison to events at the boundaries. These
timings characterize the motion of the impactor and mass shock in the
suspension, and include $\tau_d$, the time at which the impactor
reaches its maximum depth in the suspension before sinking, and
$\tau_b$, the time at which the mass shock is formed. In addition,
there is a pressure front, separate from the mass shock, that
transmits the majority of the impactor momentum to the suspension
boundary.

The propagation of stresses to the boundary provides key insight into
the these events.
Figure~\ref{fig:photoelastic-time-intensity}-A-inset shows a typical
photoelastic image from the boundary of the suspension during an
impact. In the main panel, we show the total intensity of the signal
from the photoelastic boundary over the course of the impact. Two
events are clearly visible, marked by red and yellow lines: the time
at which the first signal from the impactor reaches the
boundary,~$\tau_{p_s}$, and the time at which the intensity signal of
the boundary is maximal,~$\tau_{p_{max}}$ respectively. 120 Hz
oscillations are present in the signal due to the flickering of the
light source. These have been substantially reduced by a notch
filter. In \bl{B}, the times for the two events of part \bl{A} are
plotted vs. initial impactor speed~$v_0$, along with fitted
curves.  For a more complete picture of events in the suspension during
impact, we also plot~$\tau_a$ on the same axes. The timing of
both~$\tau_{p_s}$ and~$\tau_{p_{max}}$ is inversely proportional to
the impactor speed, and thus also inversely proportional to the
deformation rate~$v_0/D$ of the suspension, where~$D$ is the impactor
diameter, suggesting that the timescales in the suspension result from
a wave speed in the suspension which scales approximately linearly
with~$v_0$.

We gain further insight into the nature of these signals by
correlating events in the boundary with the impactor dynamics and the
bulk suspension motion. We show data in
Fig.~\ref{fig:impactor-photoelastic-timeseries}. Part A
shows~$\tau_{b}$ and~$\tau_{p_s}$ on the same axes: the time at which
the mass shock in the suspension begins moving and the time when the
first signal reaches the suspension boundary are indistinguishable,
despite the fact that the mass shock wavefront is $\sim$ millimeters below the
impactor, while the boundary is $\sim$ centimeters from the
impactor. This points to a fast timescale: information about the
impactor reaches the boundaries faster than the formation of the solid
mass beneath the impactor. We propose that this information is carried
to the boundaries by a pressure wave in the suspension, which may also
carry the bulk of the intruder momentum to the
apparatus. Figure~\ref{fig:impactor-photoelastic-timeseries}-B
compares~$\tau_{p_{max}}$ and~$\tau_{d}$: the time at which the stress
on the boundary is maximal is indistinguishable from the time when the
depth of the impactor is maximal. The very short time delay
between~$\tau_d$ and~$\tau_{p_{max}}$ further supports our argument
that there is a fast timescale for force/pressure propagation between
the suspension boundary and the impactor.

\begin{figure}
\centering \includegraphics[width=0.8\columnwidth]{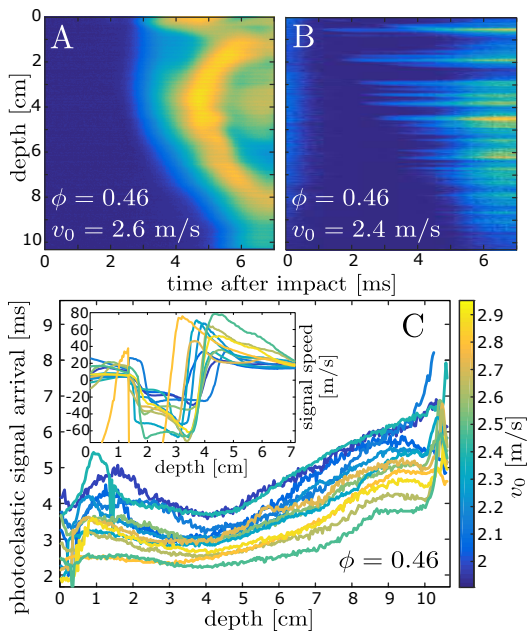}
\caption{(color online) Propagation of the pressure wave through the
  boundary of the photoelastic material. A: Space-time plot of signals
  from the edge of the suspension after impact. Blue corresponds to
  low signal intensity, and yellow to high signal intensity. B:
  Space-time plot of differences between successive frames in direct
  high speed video of the mass shock. Again blue corresponds to low
  difference and yellow to high. The stripes correspond to the
    motion of individual tracer particles. C: First arrival time of
  the pressure wave as a function of depth, for different initial
  impactor velocities~$v_0$. The pressure wave arrives first at a
  nonzero depth, then propagates both upwards and downwards along the
  suspension boundary. For clarity, individual data points have been
  joined to form lines. Inset: Speed of the pressure wave as a
  function of depth, as~$v_0$ is varied. The maximum speed of the
  pressure wave shows some dependence on~$v_0$. Again, individual data
  points have been joined to form lines. }
\label{fig:photoelastic-arrival-time}
\end{figure}

To better characterize the propagation of this pressure wave through
the suspension, we find the first arrival time of signals from the
suspension along the edge of the photoelastic boundary (see blue
dashed line in the inset of
Fig.~\ref{fig:photoelastic-time-intensity}A).
Fig.~\ref{fig:photoelastic-arrival-time}A gives a typical space-time
plot of the pressure signal moving along the boundary of the
photoelastic material. This pressure wave arrives first at a depth
of~$\approx 0.04$ m, then propagates in both directions along the
boundary away from the point where the signal first arrived, slowing
as it progresses. Fig.~\ref{fig:photoelastic-arrival-time}C shows the
first arrival times of the pressure wave along the photoelastic
material as a function of depth, for various initial impactor
speeds~$v_0$. As~$v_0$ is increased, the overall nonlinearity in the
propagation of the signal along the boundary remains consistent, in
particular in the first arrival of the wave at a depth~$\approx
0.04$m.  Fig.~\ref{fig:photoelastic-arrival-time}C inset shows the
speed of the pressure wave plotted against its depth: the speed of the
pressure wave depends strongly on its depth in the suspension along
the gel boundary.  A space-time plot of the mass shock, shown in
Fig.~\ref{fig:photoelastic-arrival-time}B, does not show a similar
nonlinearity in propagation along the boundary as the pressure wave,
supporting our argument that the pressure wave and mass shock are two
separate fronts propagating through the suspension.

\begin{figure}
\centering \includegraphics[width=0.77\columnwidth]{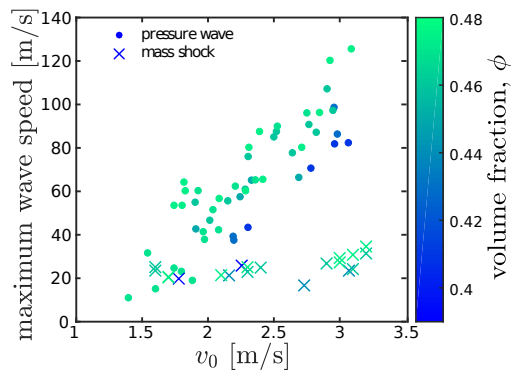}
\caption{(color online) Maximum speed of pressure wave plotted as a
  function of initial impactor speed,~$v_0$ (circles), compared to the
  maximum speed of the mass shock within the suspension
  (crosses). Data from suspension volume fractions ranging
  from~$\phi=0.39$ to~$0.475$ are shown. The speed of the pressure wave shows a systematic increase with increasing~$\phi$, while the speed of the mass shock does not.}
\label{fig:tracer-photoelastic-maxv}
\end{figure}

Figure~\ref{fig:tracer-photoelastic-maxv} explicitly compares the
maximum speed of the mass shock and the maximum speed of the pressure
wave, for different~$v_0$. We also show data for different suspension
packing fractions~$\phi$.  While the speed of the mass shock does not
show measurable dependence on~$\phi$ for the $\phi$'s studied here,
there is a systematic moderate increase in the speed of the pressure
wave for increasing~$\phi$. In addition, the mass shock and pressure
wave show different dependencies on~$v_0$, with the speed of the
pressure wave being equal to or exceeding the speed of the mass
shock. Again, this supports our argument that there exists a pressure front
in the suspension, separate from the mass shock, which precedes it and
which may propagate forces to the suspension boundary before the
arrival of the mass shock. The pressure wave speeds observed here,
which are $\sim 10^2 m/s$ for frequencies $\sim kHz$, differ
substantially from ultrasound data~\cite{Johnson13,han17} for the sound
speed, e.g. sounds speeds of $\sim 1.7 \times 10^3 m/s$ at frequencies
in the $MHz$ range. Our imaging speed of $42,000$ frames per second should be able to
detect waves at this speed, if they were present. That is, a wave of
speed $1.7 \times 10^3 m/s$ would cover the  closest
distance from intruder to side wall ($\sim 0.04m$) in $23.5 \mu s$. This is only
slightly different from our temporal resolution of $1/(4.20 \times 10^4 s^{-1}) = 23.8 \mu s$.

As noted above, we carried out additional experiments in which the
Plexiglas faces were isolated from the suspension by layers of soft
gel. These data, presented in more detail in the supplementary
materials, agree with the results of
Figure~\ref{fig:tracer-photoelastic-maxv}. 

It is also interesting to
compare the present results to experiments involving impacts into dry
granular materials~\cite{Clark15,Clark_pre16}. In these studies, the
mass flow tracked the intruder speed very closely. However, the stress
signal propagated faster than the intruder speed, and depended
nonlinearly on that speed. In the granular case, the force propagation
was known experimentally at the particle scale, which is not the case
here. Hence, comparisons between the present experiments and granular
impacts can only be qualitative.

To conclude, we observe that two separate fronts reach the boundary of
the suspension. The first is a mass shock, consistent with the front
observed in \cite{Jaeger12,Jaeger14,Waitukaitis14}. However, the total
momentum of the suspension during impact does not approach the initial
impactor momentum, suggesting that the majority of the impactor
momentum is propagated to the suspension boundary by a different
process. At the same time, information concerning the impactor
dynamics reaches the boundary of the suspension before it is carried
outward via the mass shock. That is, we observe a second front, by
visualizing the arrival of a pressure wave along our photoelastic
boundary material. The dynamics of this front are not strongly
correlated with the motion of cornstarch particles in the suspension,
but rather with the impactor dynamics. The pressure front speed,
  which grows strongly with $v_0$, and exhibits nonlinear dynamics
  along the boundary of the suspension, is generally faster than the
  mass shock speed, which grows only moderately with $v_0$. The
  pressure wave speeds observed here for timescales $\sim 1 ms$, hence
  frequencies $\sim kHz$, are an order of magnitude or more lower than
  ultrasound speeds obtained at $MHz$ frequencies~\cite{Johnson13,han17}. In addition to
  being at lower frequencies than the ultrasound measurements, the
  present experiments are in a manifestly nonlinear regime. These
  differences point to intriguing and little investigated phenomena in
  the response of cornstarch suspensions to compressive strain as a
  function of frequency and amplitude. Given the complex response of
  these suspensions to shear strain, it is not surprising that they
  also have a complex frequency and amplitude response to compressive
  strains. We close by noting a possible heuristic connection to the
  fact that one can run but not walk across a large container of
  cornstarch without sinking. Although it may be circumstantial, it is
  interesting that the pressure wave speeds observed here rise above
  the mass shock speeds for $v_0$'s that separate walking and running
  speeds.

Acknowledgements: This work is supported by NSF grant DMR1206351 and
NASA grants NNX10AU01G and NNX15AD38G.
\bibliographystyle{unsrt}
\bibliography{references}

\section*{Description of supplementary material}

\subsection*{PhotoelasticImpact.mpeg}
An impact on a suspension with~$\phi=0.47$, and~$v_0=3.1$m/s, filmed at 10,000 frames per second. The movie is played at 7 frames per second. In this experiment, the suspension was placed between crossed circular polarizers, and lit from behind, providing visual access to the strain in the photoelastic boundary.

\subsection*{TracerImpact.mpeg}
An impact on a suspension with~$\phi=0.42$ and $v_0=2.2$m/s, filmed at 10,000 frames per second. The movie is played at 30 frames per second. In this experiment, the suspension was mixed with black tracer particles, and filmed with front lighting without crossed polarizers. The resulting movie reveals the motion of the suspension during the impact. 

\subsection*{PivImpact.mpeg}

The same impact as in TracerImpact.mpeg, overlaid with the local velocity field at each point in the suspension (represented by green arrows). The local velocity field was extracted using particle image velocimetry. 

\subsection*{Coupling between Plexiglas faces and cornstarch suspension}
\begin{figure}
\includegraphics[width=0.7\columnwidth]{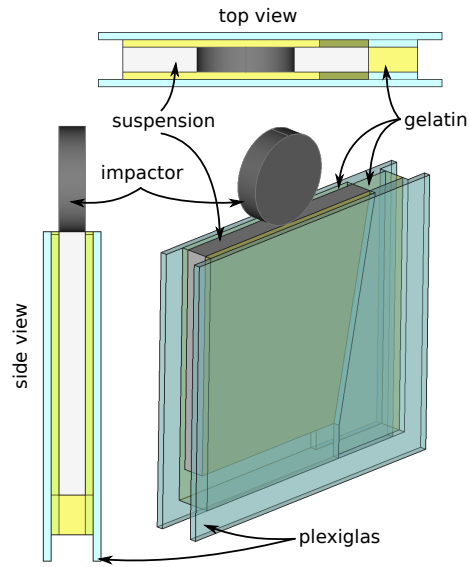}
\caption{Schematic of the experimental setup with additional layers of gelatin between the Plexiglas and cornstarch suspension. There are additional layers of gelatin in front of and behind the suspension. The gelatin is transparent, so we are able to carry out the same measurements as described in the main text. We also added a bottom boundary to the cornstarch suspension. Dark yellow in the top view indicates the sloping wall of the gelatin boundary.}
\label{fig:setup_gelatinwall}
\end{figure}

In order to test whether there was significant coupling between the Plexiglas faces of the experimental setup and the cornstarch suspension, we carried out a limited study in which additional layers of gelatin were placed between the Plexiglas and the cornstarch suspension. A schematic is shown in Fig.~\ref{fig:setup_gelatinwall}. 

\begin{figure}
\centering
\includegraphics[width=0.75\columnwidth]{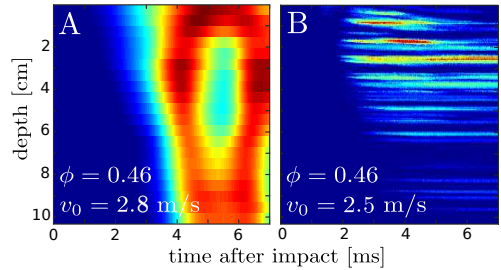}
\caption{Propagation of pressure wave and mass shock through the suspension and boundary, with additional gelatin between the Plexiglas and cornstarch suspension. A: Space-time plot of signals from the edge of the suspension after impact. Blue corresponds to low signal intensity, and red to high signal intensity. B: Space-time plot of differences between successive frames in direct high speed video of the mass  shock. Again blue corresponds to low difference and red to high difference. Compare to Fig. 4 in the main text.}
\label{fig:space-time-gelatinwall}
\end{figure}

Figure~\ref{fig:space-time-gelatinwall} shows the propagation of the pressure wave (A) and mass shock (B) in the altered experimental setup. The space time plots do not show significant qualitative differences to those shown in Fig. 4 of the main text, despite the addition of the gelatin walls. 

\begin{figure}
\centering \includegraphics[width=0.7\columnwidth]{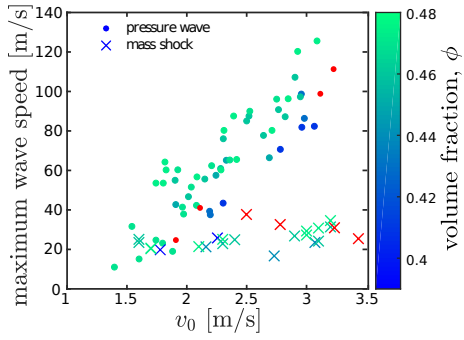}
\caption{Maximum speed of pressure wave plotted as a function of initial impactor speed,~$v_0$ (circles), compared to the maximum speed of the mass shock within the suspension (crosses). Also shown are the maximum speed of the mass shock (red crosses) and the maximum speed of the pressure wave (red circles) in the experimental setup with additional gelatin walls (red crosses). Experiments in the setup with additional gelatin walls were carried out with suspension of packing fraction~$\phi=0.46$.}
\label{fig:compare-gel-nogel}
\end{figure}

Figure~\ref{fig:compare-gel-nogel} quantitatively compares the experiments with direct contact between the Plexiglas and suspension, and the experiments with a gel layer between the Plexiglas and suspension. We overlay data for the mass shock speed with gelatin walls (red) with the original data in the main text (see Fig. 5). Within the scatter in the data, the new experiments do not produce obviously different results to the experiments without the gel layer in either the speed of the mass shock or the speed of the pressure wave, implying that the original data are not affected by the presence of the Plexiglas walls. 

\subsection*{Dependencies of impactor kinematics on~$v_0$ }

\begin{figure}
\centering \includegraphics[width=0.8\columnwidth]{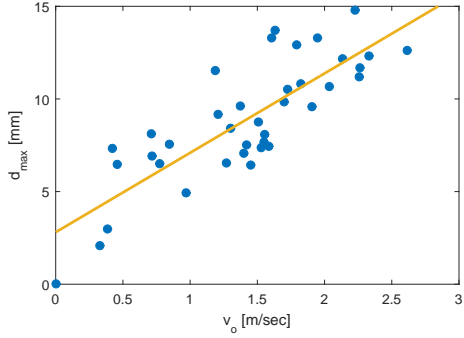}
\caption{Maximum impactor depth as a function of initial impactor velocity. Yellow indicates a linear fit.}
\label{fig:dmax}
\end{figure}

\begin{figure}
\centering \includegraphics[width=0.8\columnwidth]{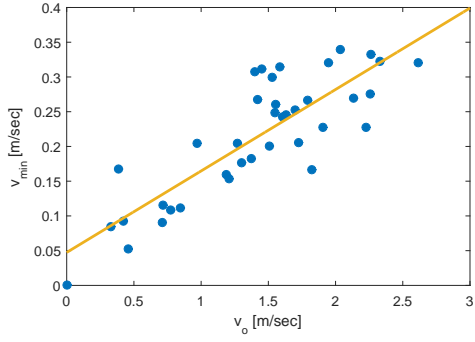}
\caption{Maximum impactor rebound velocity as a function of initial impactor velocity. Yellow indicates a linear fit.}
\label{fig:vmin}
\end{figure}

\begin{figure}
\centering \includegraphics[width=0.8\columnwidth]{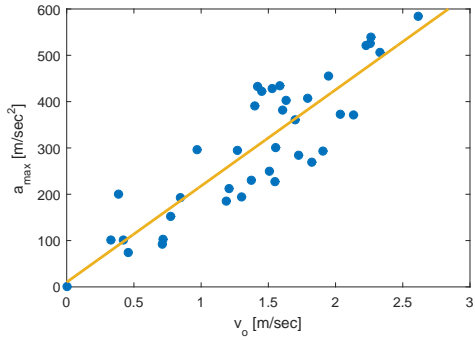}
\caption{Maximum impactor deceleration as a function of initial impactor velocity. Yellow indicates a linear fit.}
\label{fig:amax}
\end{figure}

\begin{figure}
\centering \includegraphics[width=0.8\columnwidth]{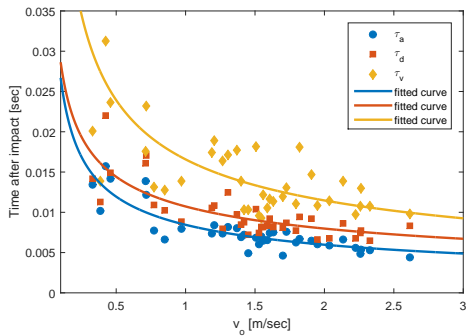}
\caption{Sequence of events in the impactor dynamics, plotted as a function
of initial impactor speed,~$v_o$. Blue dots correspond to the time at which the
deceleration of the impactor is maximal,~$\tau_a$; red squares to the time at which
the impactor reaches its maximum depth,~$\tau_d$, and yellow diamonds to the
time at which the impactor reaches its maximum rebound velocity,~$\tau_v$.}
\label{fig:timeseries}
\end{figure}

Figure 1 in the main text shows that there are several key events in the motion of the impactor during impact on a suspension:~$d_{max}$, the maximal impactor depth, ~$v_{min}$, the maximum velocity at which the impactor rebounds from the suspension, and ~$a_{max}$, the maximum impactor acceleration. We find that the magnitude of each of these events scales linearly with~$v_0$, the initial impactor velocity. Data is shown in Figs.~\ref{fig:dmax}-\ref{fig:amax}. We show also a time-series of these events, again as a function of~$v_0$, in Fig.~\ref{fig:timeseries}.

\end{document}